\documentclass{mem}
\usepackage{natbib}\usepackage{txfonts}\usepackage{balance}
\usepackage{graphicx}
\usepackage[a4paper]{hyperref}
\idline{75}{282}
\begin{document}

\title{
Variable stars in the globular cluster NGC\,2419
}

   \subtitle{}

\author{
C. \,Greco\inst{1,2}, V. \,Ripepi\inst{3}, L. \,Federici\inst{1}, G. \,Clementini\inst{1}, L. \,Di Fabrizio\inst{4}, L. \,Baldacci\inst{1}, M. \,Maio\inst{1},
 M. \,Marconi\inst{3}, I. \,Musella\inst{3}
\and P. B. \, Stetson\inst{5}
          }

  \offprints{C. Greco}

\institute{
INAF --
Osservatorio Astronomico di Bologna, Via Ranzani 1,
I-40127 Bologna, Italy
\and
Universit\`a degli Studi di Bologna,
 Dipartimento di Astronomia, Via Ranzani 1,
I-40127 Bologna, Italy
\and
INAF --
Osservatorio Astronomico di Napoli,
Salita Moiariello, 16
I-80131 - Napoli, Italy
\and
INAF --
Telescopio Nazionale Galileo, 38700,
Santa Cruz de la Palma, Spain
\and
DAO, Herzberg Institute of Astrophysics,
National Research Council,
5071 West Saanich Road,
Victoria, British Columbia, V9E2E7, Canada\\
\email{claudia.greco@bo.astro.it}
}

\authorrunning{Greco et al.}

\titlerunning{Variable stars in the globular cluster NGC\,2419}

\abstract{
We have used DOLORES at the TNG to obtain $B,V$ time series photometry of NGC\,2419, one of the 
most distant and bright clusters in the Galactic halo. These data will be used to
 study its variable star 
population in order to check whether the cluster could be the relic of an extragalactic system accreted 
by the Milky Way. Using the Image Subtraction technique 
we have identified about 300 candidate variables, many of which are in the cluster 
central regions. 
Several of the variables appear to be RR Lyrae stars,
but we detected variability also around the tip of the red giant branch, and in other regions 
of the colour-magnitude diagram. To improve the light curve sampling and to resolve variables in 
the cluster inner regions, the TNG data were combined with
HST archive data. Preliminary results are presented on the light curves from the combined data set. 
}
\maketitle{}

\section{Introduction}

NGC\,2419 is one of the most distant globular clusters (GCs) in the 
Milky Way 
(R$_{gc}\sim 90 Kpc$),  but  is much more luminous than the other 
outer-halo Galactic GCs.  In fact,  with an absolute visual magnitude M$_{V}$= $-$9.6 mag 
\citep{h96},  NGC\,2419 is among the five brightest clusters in the Galaxy.
On the other hand, NGC\,2419 cannot be considered an inner-halo cluster migrated out on an 
elliptical orbit, since its dynamical parameters  (core radius:  r$_c\sim 9 pc$, and 
half-mass radius r$_h\sim 19 pc$) are typical of an outer-halo cluster.
According to the metallicity ($\left[{\rm Fe/H}\right]  \simeq -2.12$; \citeauthor{h96} 
\citeyear{h96}) and the horizontal branch (HB) morphology,  well 
 populated from the red to the blue, NGC\,2419 belongs to the most metal-poor group of
  known Galactic GCs. \citet{h97} found that NGC\,2419 and M 92 have the same age within
 1 Gyr. 

Previous studies on the variable star content in NGC\,2419 date back to the work by 
\citet{p77} who collected photographic plates and performed
"by eye" photometry. They discovered a modest population of variable stars 
(24 {\it ab-}, 6 {\it c-}type RR Lyrae stars and  one type II Cepheid) 
in the cluster external regions. The average period of  the {\it ab-}type RR
Lyrae ($\langle P_{ab} \rangle = 0.654 \ d$) qualifies NGC\,2419 as an Oosterhoff type II \citep{o39} 
cluster.

\section{Our project: some preliminary results}
We have undertaken a new study of the variable star population of NGC\,2419 
based on accurate, deep, and high spatial  resolution  CCD photometry
covering the entire cluster.
$B,V$ time-series photometry has been obtained with DOLORES
 at the TNG. The large field of view of DOLORES ($ 9.2^{\prime} \times 9.2^{\prime} $) 
 allowed us to observe on a single shot all the known cluster RR Lyrae stars and the large 
 majority of the light from NGC\,2419.
  
Variable stars have been  identified 
using the Image Subtraction technique 
provided by the package ISIS 2.1 \citep{a00}. 

We found 283 candidate variables.

\begin{figure}[hpbt]
\begin{center}
\includegraphics[scale = 0.18]{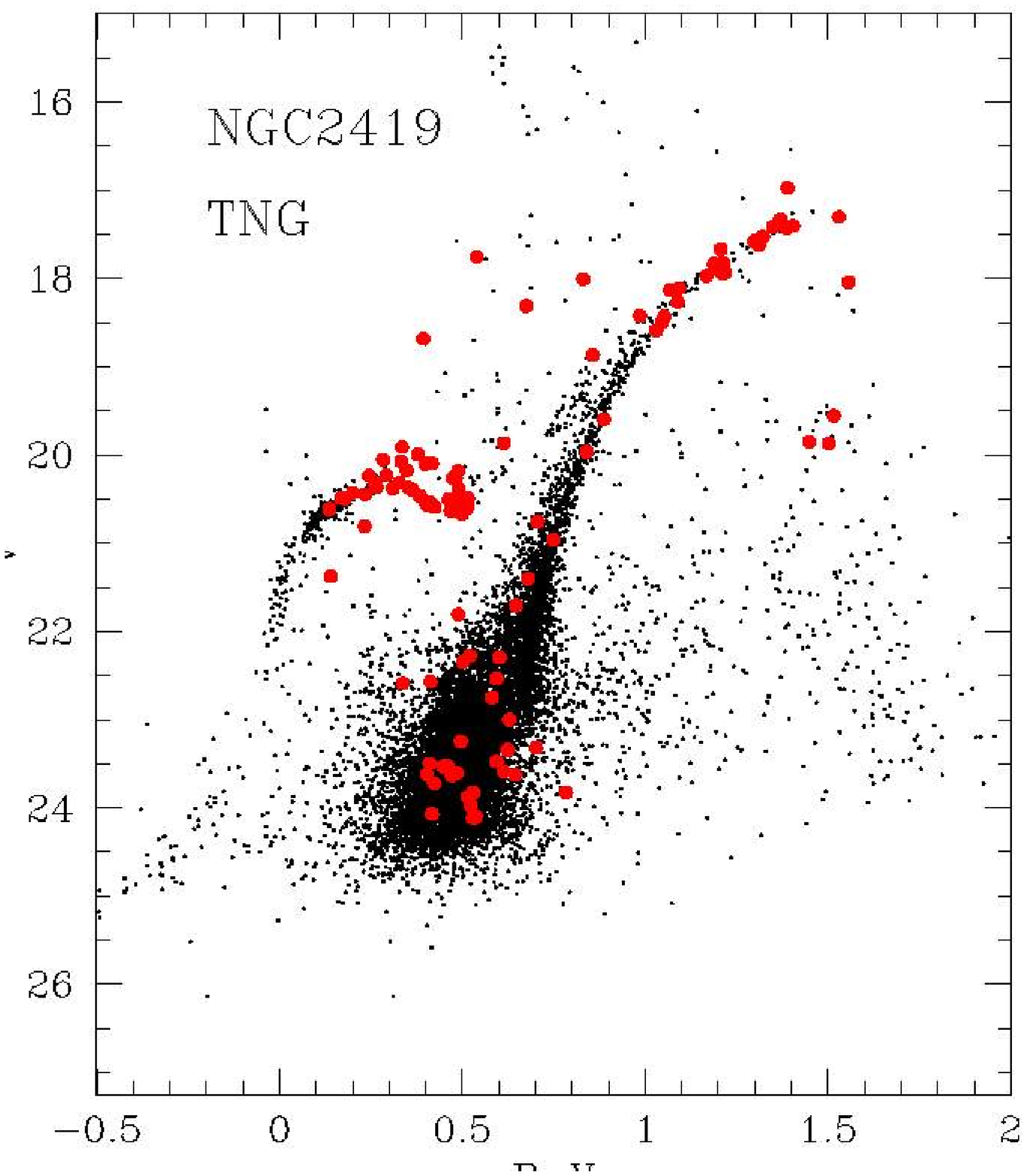}
\includegraphics[scale = 0.18]{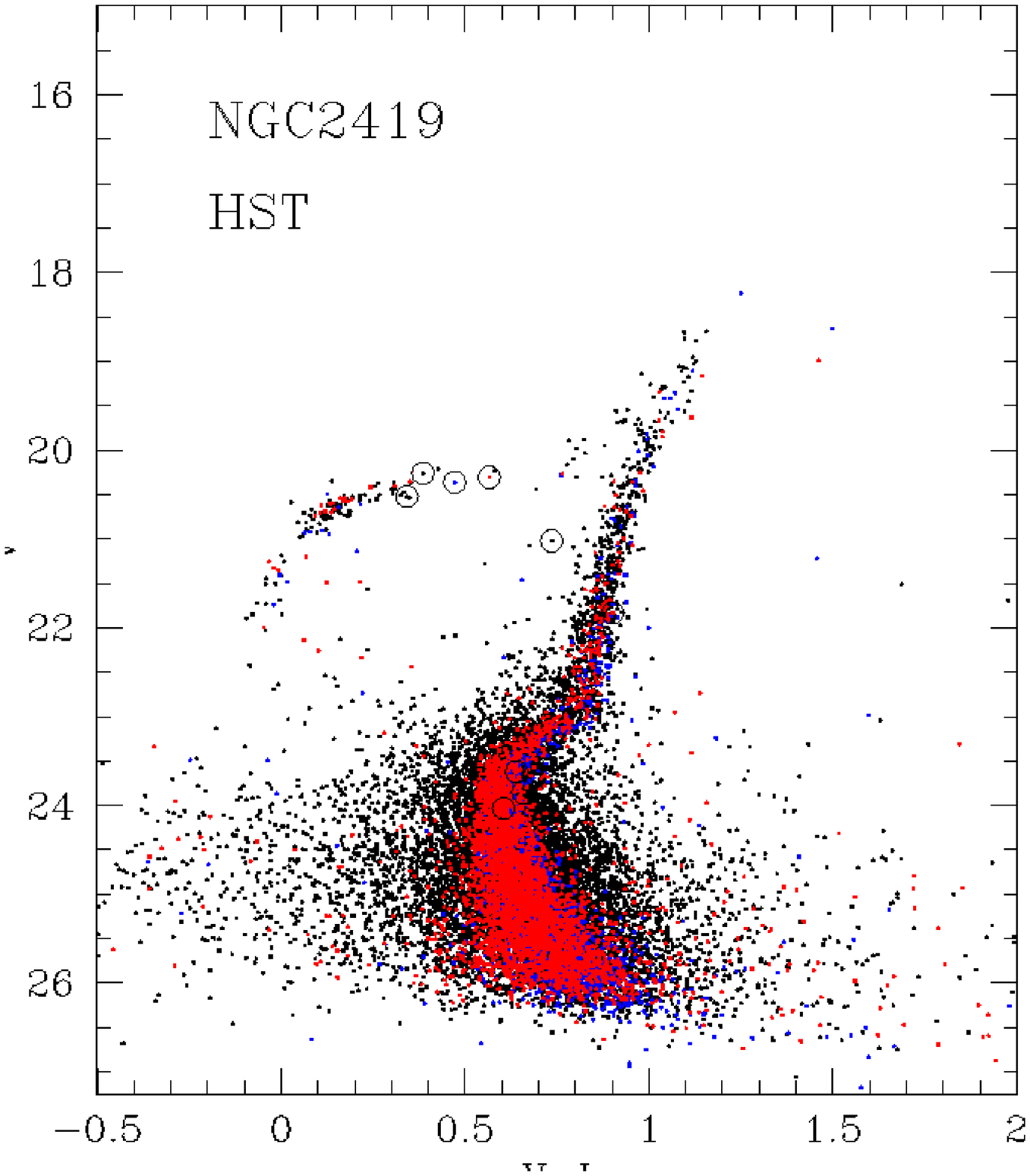}
\caption{\small{Calibrated colour-magnitude diagrams of NGC\,2419, using TNG (left panel) 
and HST-WFPC2 data (right panel).}}\label{fig:lc} 
\end{center}
\end{figure}

To extend our study  deeper inside in
the cluster centre, and to obtain  a better 
definition of the periods and light curves,  
the TNG data are being complemented with HST-WFPC2 F555W(V), F814W(I) and INT archive data.
Homogeneous photometric reductions of the entire TNG+HST+INT  data set are being performed 
with DAOPHOT-ALLSTAR-ALLFRAME 
(\citeauthor{s94} \citeyear{s94}, \citeyear{s96}).
The left panel of Figure 1 shows the colour-magnitude diagram (CMD) of NGC\,2419 from the 
TNG data. Candidate
variables are marked by large filled circles (in magenta in the electronic edition of the
journal). The cluster CMD from the HST-WFPC2 data is shown in the right panel of
Figure 1. Finally, in Figure 2 
we provide examples of light curves for variable stars
we detected in NGC\,2419.
\begin{figure}[hpbt]
\begin{center}
\includegraphics[scale = 0.20]{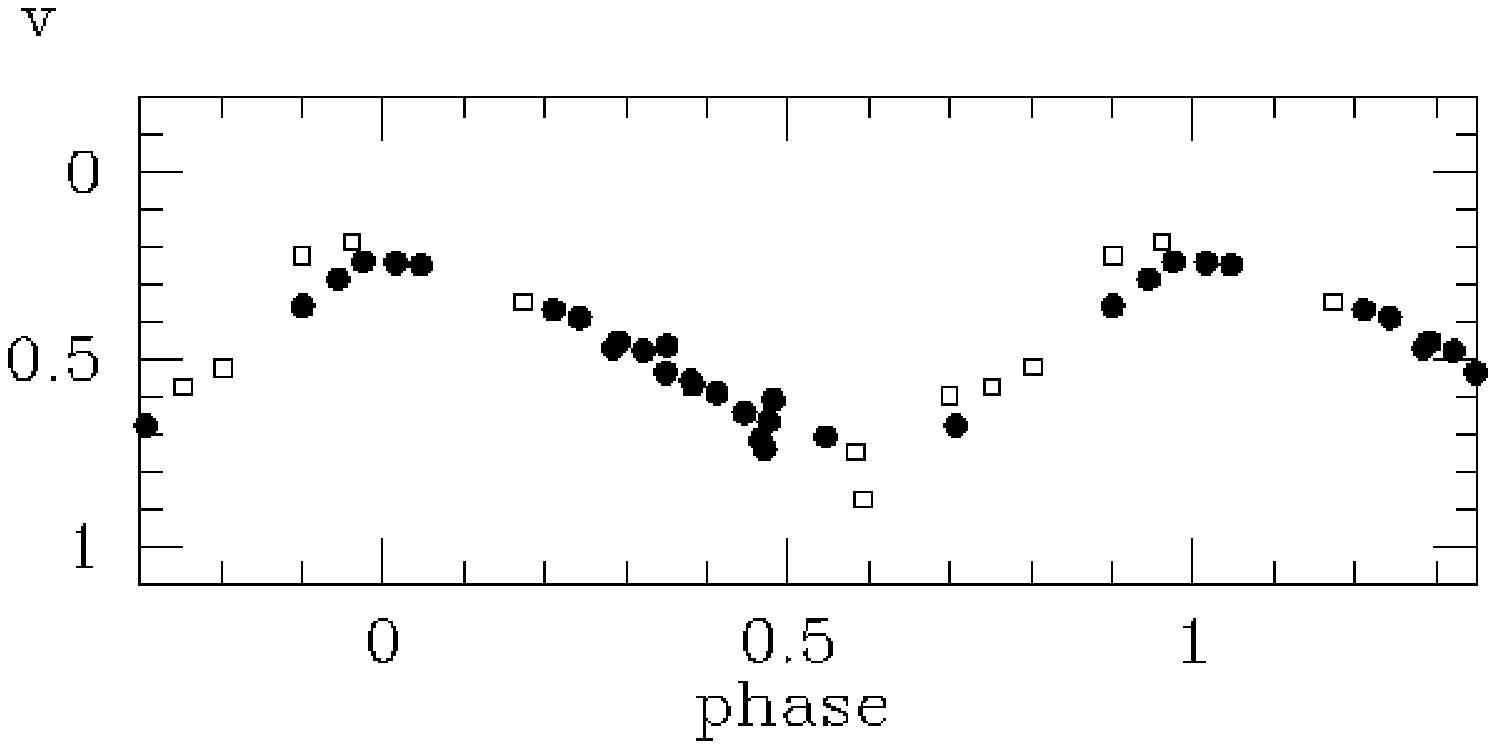}
\includegraphics[scale = 0.20]{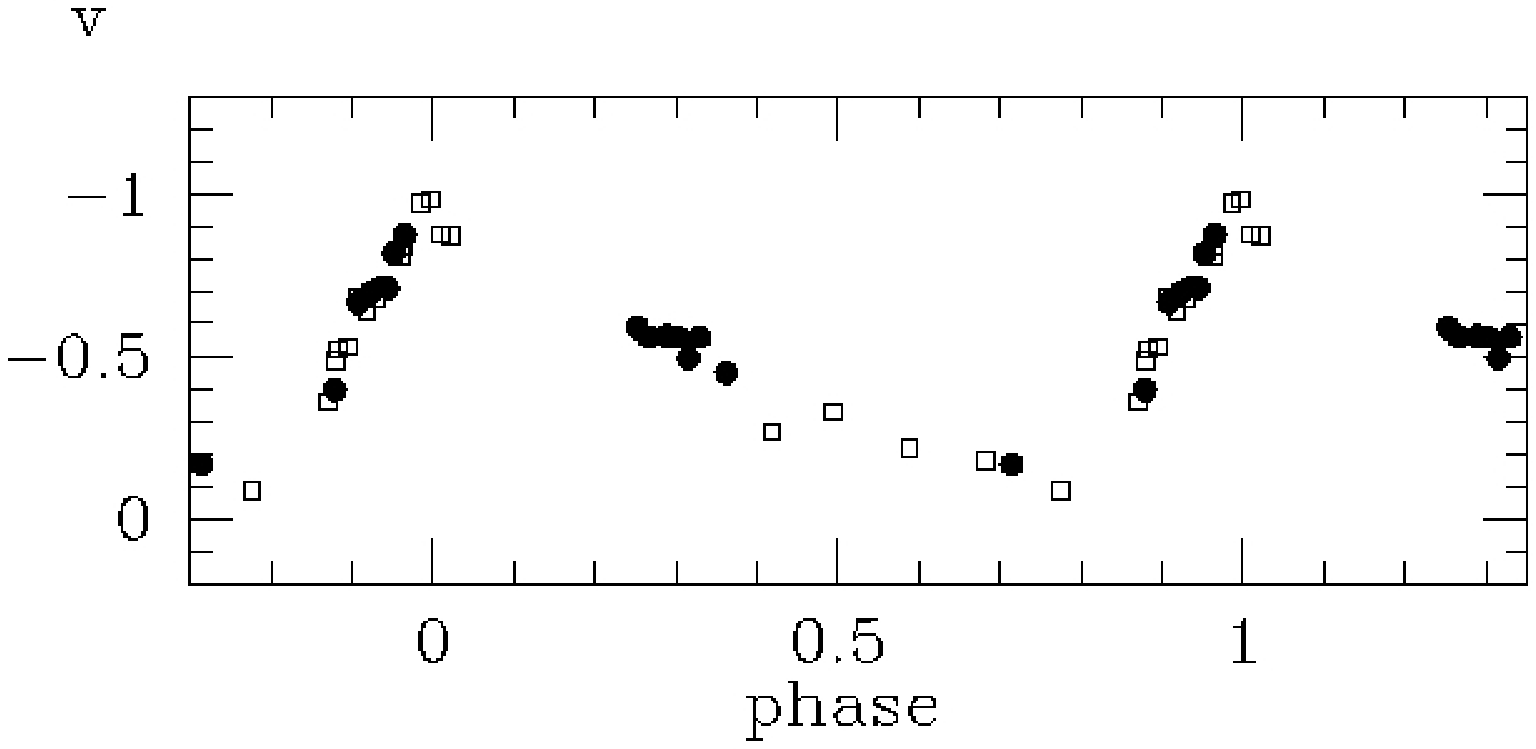}
\includegraphics[scale = 0.20]{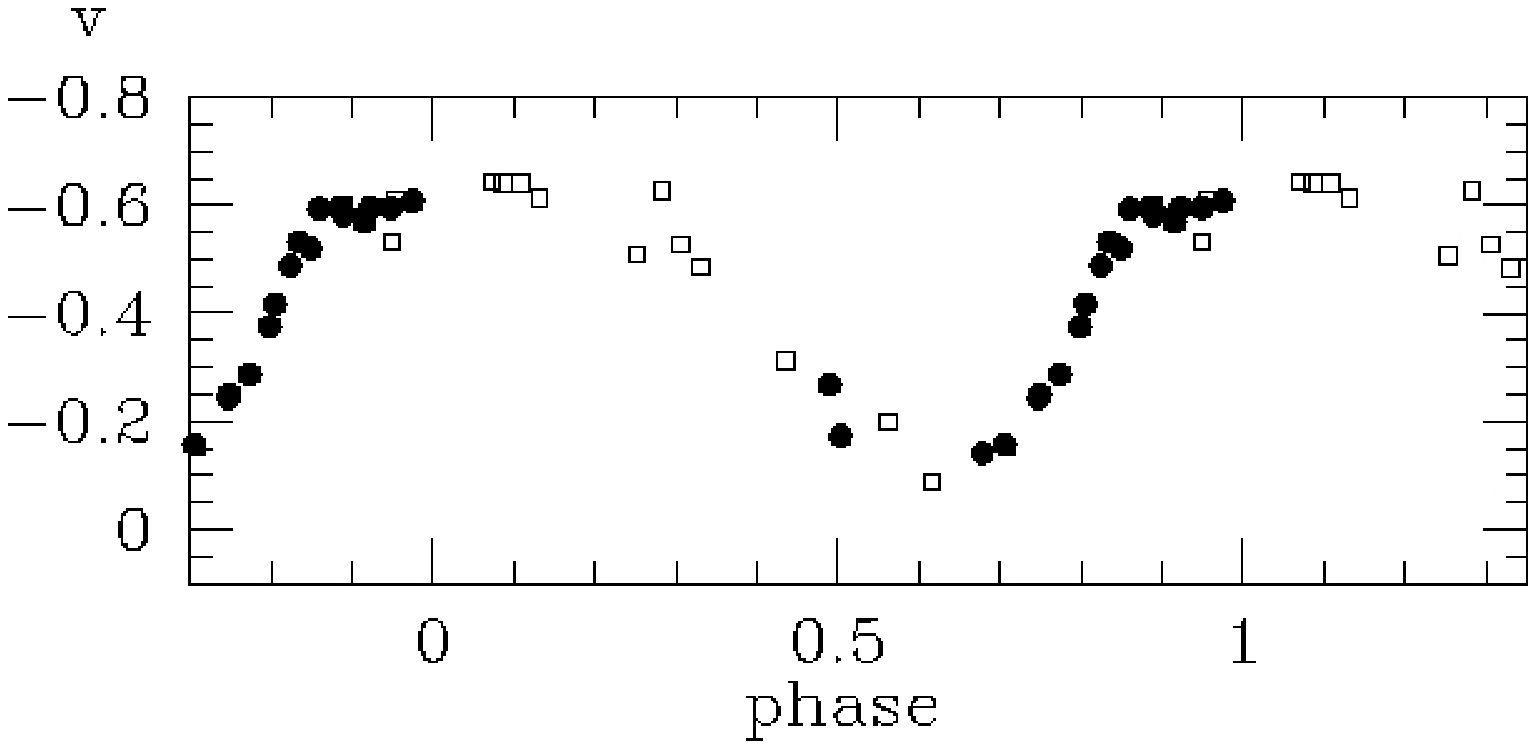}
\caption{\small{Examples of light curves for a {\it c-}, an {\it ab-}type RR
Lyrae star and 
a binary system in NGC\,2419. Filled and open symbols are the TNG and the HST-WFPC2 data,
respectively.}}
\label{fig:lc} 
\end{center}
\end{figure}
Our goals are to unveil the complete variable star population of NGC\,2419  and fully
characterize 
its pulsation properties,  
namely: light curves, periods, luminosities, amplitudes, etc.
  The comparison of both pulsation and evolutionary properties of NGC\,2419
    with those of Galactic Oosterhoff type II clusters and dwarf galaxies in 
  the Local Group will provide hints on the nature of this cluster by
  investigating the possibility that this cluster could  be the relic of the interaction 
of a vanished 
dwarf spheroidal galaxy and the Milky Way.



\bibliographystyle{aa}

\end{document}